\documentclass[journal]{IEEEtran}

\usepackage{titlesec}
\usepackage{comment}
\usepackage{amsmath}
\usepackage{subfigure}
\usepackage{graphicx}
\usepackage{subcaption} 
\usepackage{cite}
\usepackage{url}
\usepackage{color}
\usepackage{tikz}
\usetikzlibrary{arrows.meta, positioning}
\usepackage{algorithm}
\usepackage{algpseudocode}
\usepackage{amssymb}
\usepackage{flushend}
\usepackage{algorithm}
\usepackage{algpseudocode}
\usepackage{booktabs} 
\usepackage{multirow}
\usepackage{tabularx}
\usepackage{multirow}

\usepackage{graphicx}
\usepackage{subcaption}
\usepackage{tikz}
\ifCLASSINFOpdf
\else
\fi
\allowdisplaybreaks[1]
\renewcommand{\baselinestretch}{0.95}   

\begin{document}

\title{Adaptive Line Switching for Cascade Mitigation in Power Systems via Graph Reinforcement Learning}

\author{\IEEEauthorblockN{Naeem Md Sami, Soumav Biswas, and Mia Naeini}

\IEEEauthorblockA{\textit{Department of Electrical and Computer Engineering, University of South Florida, Tampa, Florida, USA} \\
naeemmdsami@usf.edu, soumavbiswas@usf.edu, mnaeini@usf.edu}}

\setlength{\textheight}{9.25in}
\IEEEaftertitletext{\vspace{-20pt}}
\maketitle
\thispagestyle{empty}

\begin{abstract} Although cascading failures are relatively rare events in power systems, they pose significant risks and can lead to severe economic and societal consequences within a short period. Therefore, developing effective mitigation strategies to prevent cascade propagation and to minimize their impacts is essential. This paper presents a deep reinforcement learning (DRL)-based framework for adaptive transmission line switching to mitigate cascading failures. Given the important role of network topology and component interactions in cascading failure propagation in power systems, graph-based DRL approaches, including Graph-DQN, Graph-A2C, and Graph-PPO frameworks, are developed to capture and utilize the topological information of power systems. The proposed frameworks are evaluated on the IEEE 39, 57, and 118 bus systems and benchmarked against the unmitigated case, random and heuristic-based switching strategies as baselines,
as well as their corresponding graph-agnostic DRL counterparts. The results demonstrate the importance of incorporating graph information into these mitigation frameworks, leading to reduced cascade size, load loss, and network clustering. 
\end{abstract}

\begin{IEEEkeywords}
Cascading Failures, Power Systems, Reinforcement Learning, Line Switching, Graph-aware Reinforcement Learning
\end{IEEEkeywords}

\section{Introduction}
Cascading failures in power transmission networks are complex, high-impact phenomena that remain a concern in modern power systems. 
A cascading failure is a sequence of dependent outages in which few initial failures can trigger subsequent interrelated failures in the system. 
Cascading failures can result in widespread service disruptions, large-scale blackouts, and significant economic losses.
\textcolor{black}{Machine learning (ML) approaches have been widely used to address various problems related to cascading failures \cite{sami2024machine}, including both situational awareness and mitigation strategies. 
For instance, deep learning and graph-based ML methods have been applied to enhance situational awareness by supporting critical component identification and cascade vulnerability assessment \cite{fan2022critical, shahzad2021vulnerability}, cascade severity prediction \cite{sami2023machine}, and failure trajectory prediction \cite{guo2026artificial}. In addition, ML-based approaches have been explored for cascading failure mitigation through corrective control actions, such as transmission line switching, load shedding, and generator redispatch \cite{zhou2025machine,zhu2021deep,zarrabian2016reinforcement}.}

The multi-stage nature of cascade propagation and its dependence on evolving system conditions make the mitigation efforts inherently a sequential decision making problem. 
Hence, reinforcement learning provides a well-suited framework for designing effective mitigation strategies.  
Specifically, Deep Reinforcement Learning (DRL) agents can learn control policies through continuous interaction with the environment (i.e., power system) and optimize long-term performance (i.e., resilience to cascading failures) under uncertainty. Recent studies have demonstrated the effectiveness of DRL in a wide range of power system applications, including voltage regulation, load restoration, economic dispatch, and cascading failure mitigation through corrective control actions \cite{li2023deep}. However, many existing DRL-based approaches represent the power system using conventional vectorized state features, which do not effectively capture the underlying network topology and spatial dependencies strongly influencing cascade propagation. 
Graph-aware DRL approaches incorporate the underlying power-system topology into the learning process, enabling DRL agents to learn optimized control policies accounting for both the system state and the topological characteristics associated with cascading failures.

In this work, both off-policy and on-policy DRL models and their graph-aware variants are investigated to mitigate cascading failures in power systems through adaptive transmission line switching.
Among the various possible corrective control actions for cascade mitigation in power systems, including load shedding \cite{das2022mitigating}, generation redispatch \cite{dai2023risk}, controlled islanding \cite{hashemi2025quantifying}, and transmission line switching \cite{zhou2025real}, 
this work focuses on transmission line switching as the intervention mechanism to relieve transmission line overloads by redistributing power flows across the network while minimizing load service interruptions.
Deep Q-Network (DQN) as the off-policy DRL model, and Advantage Actor-Critic (A2C) and Proximal Policy Optimization (PPO) as the on-policy DRL models are considered and their graph variants, namely Graph-DQN, \textcolor{black}{Graph-A2C} and Graph-PPO, are proposed for the problem of adaptive line switching for cascade mitigation. 
The graph-aware frameworks utilize Graph Neural Networks (GNN) to learn topology-aware control policies for adaptive line switching to minimize the number of transmission line failures and grid fragmentation, while maximizing served load during cascading failures.
\textcolor{black}{The key contributions of this work are summarized below:
\begin{itemize}
\item Formulated the adaptive transmission line switching problem for cascading failure mitigation in power systems as a DRL problem in which the agent determines whether to execute a transmission line switching action or maintain the current network configuration by taking no action. The proposed framework employs a balanced multi-objective reward function that seeks to minimize transmission line failures, maximize load served, and also considers system fragmentation and transmission line overload conditions.
    \item Developed graph-aware DRL frameworks based on both off-policy and on-policy learning paradigms, including Graph-DQN, Graph-A2C, and Graph-PPO. The proposed architectures are specifically designed to capture the power system's network topology through graph representations while simultaneously incorporating aggregated global system-state information.
 \item Conducted a comprehensive evaluation of the proposed graph-aware DRL models on the IEEE 39, 57, and 118 bus systems to demonstrate the value of incorporating the system's graph information into the DRL framework for cascade mitigation. The performance of the proposed Graph-DQN, Graph-A2C, and Graph-PPO frameworks was benchmarked against baselines, including the unmitigated case, and random line switching, and heuristic-based line switching strategies, with respect to cascade size, load loss, and system clustering under identical cascading failure scenarios and operating conditions.
\end{itemize}
}

\section{Related Works}

\textcolor{black}{Recent research on cascading failure mitigation has been increasingly focused on developing intelligent control strategies for remedial decisions under evolving grid conditions. Among the ML techniques explored for cascading failure mitigation, DRL has received considerable attention due to its ability to learn effective mitigation policies directly from interactions with the power system environment. More recently, GNNs have also attracted considerable attention for power system applications owing to their ability to explicitly capture the underlying network topology and spatial dependencies.} 
Here, a brief overview of the related literature on DRL applications for cascade mitigation is presented. 

The work in \cite{zhang2024mitigating} formulated cascading failure mitigation as a Markov Decision Process (MDP) and employed \textcolor{black}{the standard PPO} to learn adaptive transmission line switching policies, demonstrating improved mitigation performance over conventional remedial strategies.  
Similarly, the authors in \cite{zhou2025real} proposed \textcolor{black}{a PPO}-based framework for real-time cascade mitigation using an influence graph representation to guide proactive line switching under \textcolor{black}{load and generation fluctuations and initual outages due to natural calamities.} 
DRL has also been applied to other emergency control problems in the cascade process including load shedding. The authors in \cite{das2022mitigating} formulated cascading failures as an MDP and developed optimal load-shedding policies that minimize the expected cumulative cascade cost, gaining better performance over fixed and random shedding strategies. Moreover, the authors in \cite{pei2023emergency} integrated \textcolor{black}{GraphSAGE with double dueling Q-network} for undervoltage load shedding, showing that topology-aware representations can improve adaptability to changing network conditions. Likewise, the authors in \cite{vu2021safe} incorporated safety constraints into DRL \textcolor{black}{using a proposed augmented random search technique} for emergency load shedding, enhancing operational security during contingency events. The work presented in \cite{GRLdispatch2022iSPEC} proposed GraphSAGE-combined PPO to obtain power dispatching recommendations in grids with changing topologies, demonstrating enhanced learning of a graph-based agent.

Despite these promising advances, the application of graph-based DRL to adaptive transmission line switching remains largely unexplored. While load shedding as the control action directly disconnects load and inevitably results in service interruption, adaptive transmission line switching seeks to mitigate cascade propagation by redistributing power flows and preserving system connectivity, making it a less disruptive corrective control strategy when effective switching actions can be identified. The effectiveness of incorporating system's topological information into DRL for adaptive transmission line switching also remains insufficiently explored, particularly across power systems of different sizes and operating conditions. Motivated by this research gap, this work explores both the traditional and graph-aware DRL frameworks for adaptive transmission line switching to enhance cascading failure mitigation, as explained in the later sections.

\section{Adaptive Line Switching Problem Formulation}

\subsection{Cascading Failure Environment Modeling}

The environment is developed based on cascading failure simulations that model line overloading mechanisms following \cite{rahnamay2014stochastic} using the DC power flow (DCPF) model. DCPF, a linear approximation of AC power flow, is used in the environment to reduce computational burden and non-convergence issues, enabling repeated operations during agentic interactions. 
Cascading failures are initiated by at least two initial line outages, considering the N–1 security criterion. Following each failure (or intentional line-switching control action), power flows are redistributed. The line experiencing the highest overload level is subsequently tripped, and this process continues iteratively until no further line failures occur in the system.
Note that for simplicity and improved failure tracking, only one line outage is considered at each cascade step, assuming that the cascade process is discretized into sufficiently small time intervals.

\subsection{\textcolor{black}{Adaptive Line Switching Problem Formulation}}
The adaptive line switching problem for cascade mitigation can be formulated as an MDP with
the four-tuple vector $(\mathcal{S}, \mathcal{A}, \mathcal{R}, \gamma)$, where $\mathcal{S}$ represents the state space of the system, $\mathcal{A}$ is the discrete action space, $\mathcal{R}$ is the \textcolor{black}{reward function, and}  $\gamma \in (0, 1]$ being the discount factor. 

At each decision step $k$, the system state is represented by the vector $\underline{\mathbf{s}_k}\in \mathcal{S}$, comprising both line-specific local features and global aggregated features.
The line-specific features include the normalized active line power flow vector $\underline{\mathbf{f}_k} \in \mathbb{R}^M$, the power injection differences at both branch ends, $\underline{\mathbf{p}_k} \in \mathbb{R}^M$, the binary status of the line $\underline{\mathbf{n}_k} \in \mathbb{R}^M$, and the cluster index for the lines $\underline{\mathbf{g}_k} \in \mathbb{R}^M$, where $M$ is the number of transmission lines in the system. The active line power flow is normalized with respect to the line capacity. Moreover, for the line $m$ connecting bus $i$ and $j$, the power injection difference is defined as, $p_m=p_i - p_j$.  The binary line-status vector, $\underline{\mathbf{n}_k} \in \mathbb{R}^M$ also captures the state of each line as $n_k^m=1$ if line $m$ is active and $n_k^m=0$, otherwise. During the cascade process due to failures or line switching actions, \textcolor{black}{multiple clusters of connected components may form}. To track \textcolor{black}{this} clustering process, an integer is assigned to each line, to indicate which cluster a particular line belongs to. The numbers are assigned with respect to the order the grid is fragmented, and all the lines in the connected component $\mathcal{C}$ receive the same index value. The vectors define the line-localized features as $\underline{\mathbf{X}_k} = [\underline{\mathbf{f}_k}; \underline{\mathbf{p}_k}; \underline{\mathbf{n}_k}; \underline{\mathbf{g}_k}]$

The global aggregated features at \textcolor{black}{time} step $k$ include the total number of clusters formed in the system, $c_{k}$, 
the total number of overloaded lines, $o_k$,
the total amount of load loss due to disconnection, $d_k$, and the binary flags: \textcolor{black}{ $l^s_{k}$, $l^c_{k}$}, and $u_{k}$ 
to indicate whether the grid configuration has been modified by line switching, whether there is a new failure due to cascade, and whether the cascade has terminated, respectively. 
Note that the states with $u_k=1$ are absorbing states. The global features can be summarized as \textcolor{black}{$\underline{\mathbf{Z}_k} = [c_k, o_k, d_k, l^s_{k}, l^c_{k}, u_k]$.}
The state of the system, including both local and global state features, is thus $\underline{\mathbf{s}_k} = [\underline{\mathbf{X}_k};\underline{\mathbf{Z}_k}]$.
The transitions among the states are governed by the underlying power system dynamics (in the simulated environment).

\textcolor{black}{The action space contains all possible single-line switching options and one no-action decision. Specifically, the action set is $\mathcal{A} = \{a^0, a^1, …, a^M\}$, where $a^0$ indicates no intentional line switching and $a^i$ indicates the action of disconnecting line $i$.}

The reward function is formulated to incentivize the agent to minimize the number of overloaded transmission lines at each time step, reduce the total number of line outages (including both control-induced switching actions and cascade-induced failures), minimize load loss across the system, and promote early cascade termination. The reward $r_{k}
= f(\underline{\mathbf{s}_k},a_k,\underline{\mathbf{s}_{k+1}})
$ at time step $k$ after action $a_k$ is defined as 

\begin{equation*}
r_{k} = - (d_{k}-d_{k+1})/D – l^s_{k+1}/{M} - l^c_{k+1}/{M} 
\end{equation*}
\begin{equation}
+ 0.075 \times (o_{k}-o_{k+1})/M + 0.5 \times u_{k+1},
\end{equation}

where $D$ is the total load in the system used to normalize the first term and the number of overloaded lines is normalized by $M$. 
The weights assigned to the individual reward components were empirically tuned through iterative experimentation to achieve a balanced trade-off among the desired control objectives.
In the proposed reward function, the first three negative reward components penalize load shedding, transmission line switching (representing the cost associated with line switching control action), and line outages caused by cascading failures, respectively. \textcolor{black}{The fourth component of this function rewards decreases in the number of overloaded lines and penalizes increases in the number of overloaded lines between consecutive states.} The final component provides an additional reward when the cascade stops and no further line failures occur in the system.

The DRL agent observes the system state, denoted by $\underline{\mathbf{s}}_k$, from the cascading failure environment and selects an action $a_k$. If the selected action is \emph{no action}, the cascade evolves naturally, resulting in a transition to a new state that may include additional line outages or a cascade termination state. Conversely, if the agent chooses to disconnect a transmission line, the resulting power flow redistribution causes the system to transition to a new state with updated system configuration, which may lead to subsequent line failures or cascade termination.

\section{Graph-Agnostic DRL Frameworks}

As discussed previously, adaptive transmission line switching for cascading failure mitigation in power systems is inherently a sequential decision-making problem. Here, the problem is formulated within a DRL framework, where, following each line outage during the cascade, the agent evaluates the available line-switching actions, including the option of taking no action, and selects the action that yields the greatest expected benefit as captured in the reward. This section presents the architecture of the graph-agnostic DRL framework.

\subsection{DQN Model}

DQN is a value-based off-policy DRL technique with the Q-learning mechanism empowered by a deep neural network. It relies on the generation of Q-values from the state-action pair $Q(s,a)$. 
Below, the building blocks for the DQN architecture are reviewed sequentially.

\textit{Line Embedding Layer Block:} 
This block processes the line-localized features for each line, denoted as $\mathbf{X}_{m,k} \in \mathbb{R}^4$ (i.e., the four localized features of line $m$ from vector $\underline{\mathbf{X}_k}$) to obtain the embedding $h^0_{m,k} = n_{m,k}\tanh(W_0\mathbf{X}_{m,k}+b_0)$, where $W_0$ and $b_0$ are the weight and bias,  and $n_{m,k}$ is the binary status of line $m$ from $\underline{\mathbf{n}_k}$. Therefore, for $M$ lines, the input line vector feature $\underline{\mathbf{X}_k} \in \mathbb{R}^{M\times 4}$ is converted into an embedding, $\underline{h^0_k} \in \mathbb{R}^{M\times d_1}$, where $d_1=128$ is the network width. 

\textit{Self-Residual Layer Block:} To preserve local information during updates and facilitate gradient propagation, two self-residual layers are employed, each with a network width of $d_1=128$, to process the output of the {\it Line Embedding Layer}. For line $m$ at time step $k$, the $j$th residual layer computes $T_{m,k}^{j} = W_{S}^{j} h_{m,k}^{j-1} + b^{j}$ and updates the hidden representation as $h_{m,k}^{j} = n_{m,k}\,\mathrm{LN}\!\left[\tanh\!\left(T_{m,k}^{j}\right) + h_{m,k}^{j-1}\right]$, where $\mathrm{LN}(\cdot)$ denotes the layer normalization operation. \textcolor{black}{For $M$ lines with two self-residual layers, the embedding is represented as $\underline{h^2_k} \in \mathbb{R}^{M\times d_1}$.}

\textit{Mean-max Pooling Block:}
In this block, the mean pooling operation processes the output of the previous layer to capture the overall system conditions across all lines and the max pooling captures the most critical localized conditions across the lines. The output of the mean-max pooling for $M$ lines is \textcolor{black}{$\underline{h_{pool}} = [\underline{h_{mean}};\underline{h_{max}}] \in \mathbb{R}^{M\times 2d_1}$.}

\textit{Transformed Global Embedding Block:} 
The global features ${\underline{\mathbf{Z}_k}}$ discussed in the previous section will be combined with the output of mean-max pooling layer as \textcolor{black}{ $[\underline{h_{pool}}; \underline{\mathbf{Z}_k}] \in \mathbb{R}^{M\times2d_1+6}$} and processed in this block into $\underline{h_{T_k}} \in \mathbb{R}^{M\times d_1}$ to generate a summary of the complete cascade state consisting of global information.

\textit{Q-value Head and No-action Head Blocks:} After obtaining the transformed global embedding $\underline{h_{T_k}}$, 
at each time step $k$, the DRL agent selects an action from the available action space, which consists of either taking no action or switching one of the active transmission line. In order to obtain the Q-values for these two classes of actions, two neural network heads, named `Q-value head' and `no-action head' are designed. The Q-value head takes the input feature $\underline{\mathbf{X}_k}$, the residual output $\underline{h^2_k}$, 
and the transformed global embedding, $\underline{h_{T_k}}$ to determine the Q-values for each of the $M$ lines. On the other hand, the no-action head takes the transformed global embedding, $\underline{h_{T_k}}$, as the only input to calculate a single no-action Q-value. 
Upon obtaining these $M+1$ Q-values, an action mask is used to filter previously failed or switched lines from the set of possible actions. The action at time $k$ is selected using the $\epsilon$-greedy mechanism as $a_k = \arg\max_{a \in \mathcal{A}} Q_{\theta}(s_k,a)$, where $\theta$ denotes the parameters of the Q-network.

\textit{DQN Value Update Block:} After executing an action, the resulting transition, consisting of the current state, action, reward, and next state, is stored in the replay buffer. Mini-batches of past transitions are then sampled to train the online Q-network, while a separate target network is used to provide stable learning targets. The network parameters are updated by minimizing the difference between the predicted and target Q-values using the Huber loss function.

The DQN is trained using the Adam optimizer with a learning rate of $3 \times 10^{-5}$ and a replay buffer capacity of $10{,}000$ transitions. An $\epsilon$-greedy exploration strategy is employed, with the exploration rate decayed every $5{,}000$ transitions. Experience replay is performed using mini-batches of size $128$.

\begin{figure*}[htbp]
    \centering
    \captionsetup[subfigure]{justification=centering}
    \begin{subfigure}
        \centering
        \includegraphics[height=5cm,width=16cm]{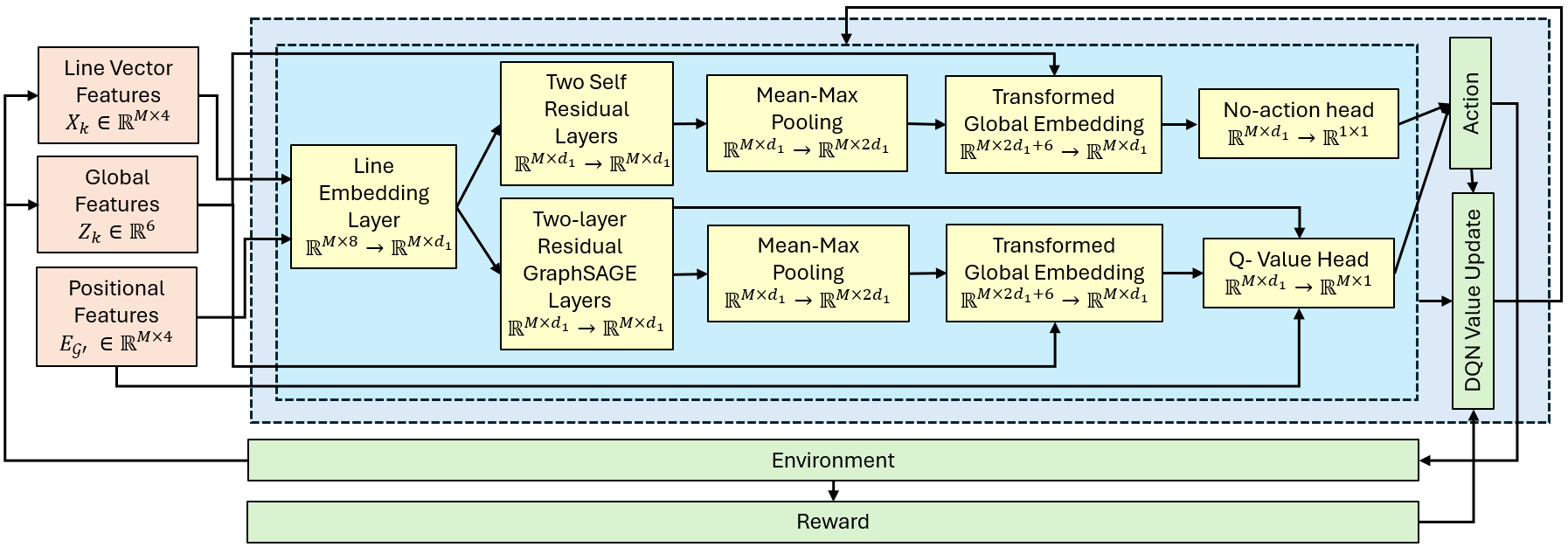}
        \caption*{(a)}
    \end{subfigure}
    \begin{subfigure}
        \centering
        \includegraphics[height=7cm,width=16cm]{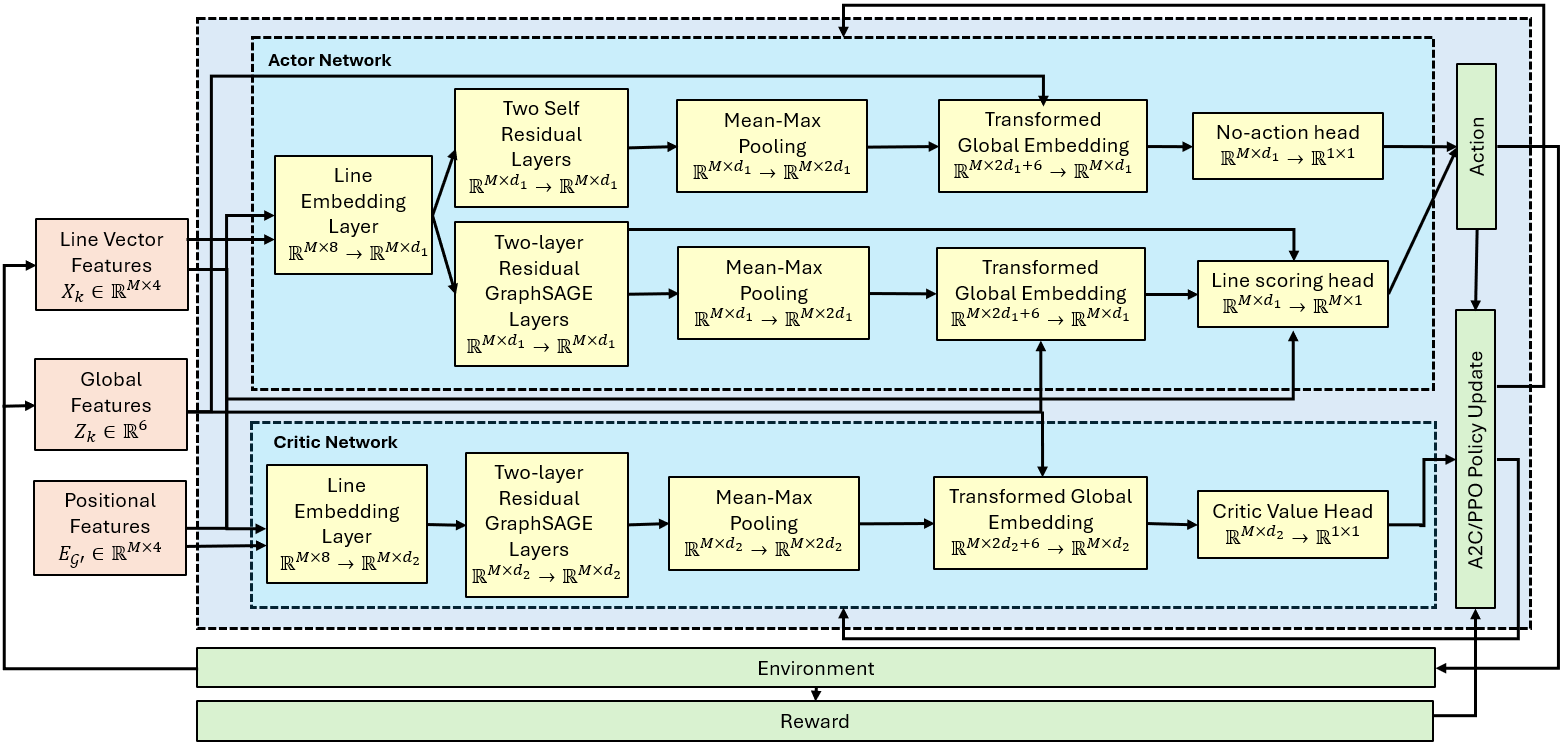}
        \caption*{(b)}
    \end{subfigure}
    \caption{General schematic of the proposed (a) Graph DQN and (b) Graph A2C/PPO techniques.}
\end{figure*}

\subsection{A2C and PPO Models} \textcolor{black}{This work employs two on-policy DRL techniques, namely A2C and PPO, to address the transmission line switching problem.}
A2C \cite{RL_A2C} is a neural network-based on-policy actor-critic algorithm that integrates value-based and policy-based learning. The actor updates the policy by favoring actions with positive advantage estimates, while the critic improves the accuracy of value predictions by minimizing the value function estimation error. PPO \cite{RL_PPO} is another advanced DRL algorithm that utilizes a clipped surrogate objective and several mini-batch epochs to achieve stable updates, unlike A2C, which considers single-step updates on the complete rollout batch. 
Both of these models are divided into two major blocks: the {\it Actor Network Block} and the {\it Critic Network Block}. The actor network learns the policy $\pi_\theta(a_k|s_k)$ with parameter $\theta$ and recommends action $a_k$ from state $s_k$ for time step $k$. The critic, on the other hand, estimates the state value $V_\phi(s_k)$ with parameter $\phi$ to predict the expected future cumulative reward from the current state. The steps in each of these blocks are elaborated below:

\emph {Actor Network Block:} Similar to the DQN model, the actor network for both A2C and PPO models takes the line-localized features $\underline{\mathbf{X}_k}\in \mathbb{R}^{M\times 4}$ as the input, feeds it into the \textit{Line Embedding Layer Block}, as discussed in Section~IV.~A, to obtain the initial embedding $\underline{h^0_k} \in \mathbb{R}^{M\times d_1}$. The obtained embeddings will pass through the \textit{Self-Residual Layer Block} with two layers, mentioned in Section~IV.~A, to obtain $\underline{h^2_k} \in \mathbb{R}^{M\times d_1}$, which then goes through the \textit{Mean-max Pooling Block}, and the \textit{Transformed Global Embedding Block} to obtain the embedding $\underline{h_T} \in \mathbb{R}^{M\times d_1}$. 
The actor network has a width of $d_1=128$. The purpose of these blocks is to generate actions based on the scores generated for each line and one score generated for the no-action. The actor has two decision-making neural network heads, named `line scoring head' and `no-action head', to calculate the probability for the actions. Instead of the Q-values in the DQN model, the actor generates $M+1$ logits ($M+1$ is the number of possible actions, including $M$ lines to switch and the no-action) from both heads, \textcolor{black}{pass them through the action mask to filter invalid lines,} and converts them into probabilities using the softmax function to pick the highest probable action for the next steps.

\emph{Critic Network Block:} The critic network also consists of the same blocks reviewed in the DQN model. 
The only difference is that for the critic network, the network width is chosen to be $d_2 = 192$, and the output from the \textit{Transformed Global Embedding Block} goes to a \textit{critic value head} to calculate a scalar critic value function, $V_{\phi}(s_k)$, to be fed into the \textit{A2C/PPO Policy Update Block}.

Once the action probabilities are produced by the \emph{Actor Network Block} and the state-value estimate is generated by the \emph{Critic Network Block}, both outputs are passed to the \emph{A2C/PPO Policy Update Block}, where the policy and value networks are updated according to the learning mechanism of the selected algorithm. The policy update procedures for A2C and PPO are briefly described below.

\textit{A2C Policy Update:} At each time step $k$, the A2C algorithm computes the temporal-difference (TD) target, $y_k$, and the corresponding advantage estimate, $A_k$, to measure if the transition is better than the critic prediction. 
\textcolor{black}{The A2C uses the one-step target directly as: $y_k =  r_{k} + \gamma(1-u_{k}) V_{\phi}(s_{k+1})$ and $A_k = y_k – V_{\phi}(s_{k})$}, 
where $r_k$ is the reward and $u_k$ is the binary termination flag for cascade step $k$. The advantage values are subsequently normalized across each training batch to improve learning stability. The actor network minimizes the loss function, $\mathcal{L}_{actor} = -\mathbb{E}[A_klog\pi_{\theta}(a_k|s_k)+\beta \mathcal{H}_k]$, and the critic network minimizes the loss function, $\mathcal{L}_{critic} = \frac{1}{2}(V_{\phi}(s_k)-y_k)^2$, where $\beta$ is the entropy coefficient, $\mathcal{H}$ is the policy entropy, and $\mathbb{E[.]}$ is the expectation operation.

\textit{PPO Policy Update:} \textcolor{black}{At each time step $k$, PPO policy update uses the difference target, $y_k$, and the advantage estimate, $A_k$, as: $y_k =  r_{k} + \gamma(1-u_{k}) V_{\phi}(s_{k+1}) - V_{\phi}(s_k)$ and $A_k = \sum_{t=0}^{K-k-1}(\gamma \lambda)^t y_{k+t}$}, 
where, $r_k$ is the reward, $u_k$ is the binary termination flag for cascade step $k$, \textcolor{black}{$K$ is the episode length}, $\gamma$ is the discount factor, $\lambda$ is the generalized advantage estimates (GAE) parameter and $V_{\phi}(s_k)$ is the value function for state $s_k$.  
A batchwise advantage normalization is performed for stable training. PPO then uses the probability ratio $\rho_k(\theta)$ with respect to the old policy $\pi_{\theta_{old}}$ and new policy $\pi_{\theta}$ as $\rho_k(\theta) = \frac{\pi_{\theta}(a_k|s_k)}{\pi_{\theta_{old}}(a_k|s_k)}$ with parameter $\theta$, 
to calculate the actor loss with clipping as $\mathcal{L}_{actor} = -\mathbb{E}[\text{min}(\rho_kA_k, \text{clip}(\rho_k, 1-\delta, 1+\delta)A_k)]$,
\textcolor{black}{where $\delta$ is the PPO clipping parameter and $\mathbb{E[.]}$ is the expectation operation.}
The critic loss with respect to the old value function $V_{\phi_{old}}(s_k)$ and new value function $V_{\phi}(s_k)$ is $\mathcal{L}_{critic} = \frac{1}{2}\mathbb{E}[(V_{\phi}(s_k) - V_{\phi_{old}}(s_k)+A_k)^2]$.

The actor and critic networks for both A2C and PPO are trained using the Adam optimizer with a learning rate of $3 \times 10^{-4}$. 
For each transition in PPO, the GAE is obtained using $\gamma = 1$ and $\lambda = 1$, followed by batchwise advantage normalization. Since PPO uses the clipped surrogate objective, a clipped interval of $[0.8, 1.2]$ is chosen with an exploration rate of $\epsilon = 0.02$. The entropy regularization changes from $0.03$ to $0.008$, uses a minibatch size of $128$, and an early optimization stopping criterion is set when the Kullback-Leibler (KL) divergence surpasses $0.02$.

\section{Graph-aware DRL Frameworks}

In this section, graph-aware extensions of the DRL models introduced in Section~IV, namely Graph-DQN, Graph-A2C, and Graph-PPO, are presented. These models are designed to capture the underlying network topology, enabling the incorporation of structural information that can play important roles in the propagation and mitigation of cascading failures in power systems.

Because the cascading failure model focuses on transmission line overloads and outages, the features used by the learning agents are primarily line-based. Accordingly, the original power system graph, $\mathcal{G}=(\mathcal{V},\mathcal{E})$, is transformed into its corresponding {\it line graph} $\mathcal{G}'=(\mathcal{V}',\mathcal{E}')$, where each node represents a transmission line in the original network, i.e., $\mathcal{V}'={e_1,e_2,\ldots,e_M}$ with $e_i\in\mathcal{E}$. Two nodes in $\mathcal{G}'$ are connected if their corresponding transmission lines in the original graph share a common endpoint (bus), thereby capturing the topological relationships among neighboring lines that influence cascading failure propagation.
This line graph is one of the inputs of the graph-aware DRL models as discussed in the following subsections. Figure 1(a) and (b) depict the general schematic for the graph-aware DQN and graph-aware A2C/PPO techniques, respectively. Next, the architectures and their respective sequential blocks are discussed in detail.

\subsection{Graph DQN Model}
The discussion of the graph-aware DQN in this subsection focuses only on the components that differ from or extend the DQN architecture presented in Section~IV. 

\textit{Positional Feature Block:} To incorporate topological information in the graph-aware models, the positional features based on the line graph adjacency matrix $\mathbf{A_{\mathcal{G'}}} \in \mathbb{R}^{M\times M}$, are introduced \cite{LPE}. 
For the matrix $\mathbf{A_{\mathcal{G'}}}$, the normalized line graph Laplacian matrix $\mathcal{L}_{\mathcal{G'}}$ is derived with respect to its degree matrix, $D_{\mathcal{G'}}$ and identity matrix $I$ as $\mathcal{L}_{\mathcal{G'}}=I-(D_{\mathcal{G'}}^{-1/2}\mathbf{A_{\mathcal{G'}}}D_{\mathcal{G'}}^{-1/2})$. Upon performing the eigendecomposition of the matrix $\mathcal{L}_{\mathcal{G'}}$, the positional feature for graph $\mathcal{G'}$ is defined as
${\underline{\mathbf{E}_\mathcal{G'}}} = [\underline{\upsilon_2}; \underline{\upsilon_3}; \underline{\upsilon_4}; \underline{\upsilon_5}]\in \mathbb{R}^{M\times4}$, where $\underline{\upsilon_i}\in \mathbb{R}^M$ is the corresponding eigenvector to the $i$-th sorted eigenvalue in ascending order, to capture key topological information associated with the first four non-zero lower graph frequencies.

\textit{Line Embedding Layer Block:} 
For the graph-aware models, this block processes both the line-localized feature, $\underline{\mathbf{X}_k}$, and positional features $\underline{\mathbf{E}_\mathcal{G'}}$. For $M$ lines, \textcolor{black}{the input combined feature vector $\underline{\hat{\mathbf{X}_k}}=[\underline{\mathbf{X}_k}; \underline{\mathbf{E}_\mathcal{G'}}] \in \mathbb{R}^{M\times 8}$} is converted into an embedding, $\underline{h^0_k} \in \mathbb{R}^{M\times d_1}$, where $d_1=128$. This generated embedding goes into two separate blocks, i.e., the \textit{Residual GraphSAGE Layer Block} and the \textit{Self-residual Layer Block}, to generate \textcolor{black}{Q-values for two classes of actions:} 
the first block produces Q-values for all line-switching actions, while the second block generates a single Q-value associated with the no-action.
\textcolor{black}{The block uses the combined features $\hat{\mathbf{X}}_{m,k}\in\mathbb{R}^8$ (i.e.,
the four localized features and four positional features of line $m$ from vector $\underline{\hat{\mathbf{X}}_k}$)} to obtain the embedding $h^0_{m, k} = n_{m, k}\tanh(W_0 \hat{\mathbf{X}}_{m,k} + b_0)$, where $n_{m, k}$ is the binary status of the line $m$, and $W_0$ and $b_0$ are the weight and bias. 

\textit{Residual GraphSAGE Layer Block:} The model employs GraphSAGE \cite{RL_GraphSAGE} due to its flexible neighborhood aggregation mechanism. Unlike many GNN architectures that require processing the entire graph during each update, \textcolor{black}{GraphSAGE samples and aggregates a fixed set of active neighboring nodes, significantly reducing memory requirements and computational complexity.} This scalability and efficiency make it particularly well-suited for time-critical applications such as cascading failure mitigation in power systems, where rapid decision-making is essential.
In this model, the embedding, $\underline{h^0_k}$, obtained from the previous \textit{Line Embedding Layer Block} passes through two residual GraphSAGE layers. \textcolor{black}{Each GraphSAGE layer $j$ uses the line adjacency matrix, $\mathbf{A}_\mathcal{G'}$ to guide the aggregation. Specifically, an active line $m$ aggregates the embeddings of its active neighbors as
$ \bar{h}^{j-1}_{m,k}=\frac{\sum_{i=1}^{M}\mathbf{A}_{\mathcal{G'}_{m,i}}n^i_k h^{j-1}_{i,k}}{\sum_{i=1}^{M}\mathbf{A}_{\mathcal{G'}_{m,i}}n^i_k}$,
where $n_k^i$ is the binary status of the line $i$ at time $k$.} 
Each GraphSAGE layer $j$ combines the self transformation, $T_{S_{m,k}}^j$ and the neighborhood transformation, $T_{G_{m,k}}^j$, together with a learnable parameter $\alpha_j$ into a message $T_{C_{m, k}}^j$ using the equation $T_{C_{m, k}}^j = T_{S_{m, k}}^j +\alpha_j (T_{G_{m, k}}^j - T_{S_{m, k}}^j)$, where $T_{S_{m, k}}^j = W_{S,j}h_{m,k}^{j-1}+W_{G,j}h_{m,k}^{j-1}+b_j$ and $T_{G_{m, k}}^j = W_{S,j}h_{m,k}^{j-1}+W_{G,j}\bar{h}_{m,k}^{j-1}+b_j$, 
and $\alpha_j = \sigma(\eta_j)$, $\eta_j$ being a trainable scalar, going through the sigmoid function, $\sigma$, to limit $\alpha_j$ from 0 to 1. The weight matrices $W_S, W_G$ are for self-transformation and neighborhood transformation, respectively, and $b$ is the bias.
The combined transformation, $T^j_{C_{m,k}}$, then goes through the $\tanh$ activation function, adds a residual connection, $h^{j-1}_{m,k}$, to retain the old information, and finally performs layer normalization to have consistent activation magnitude for the aggregations. The updated embedding from the residual GraphSAGE layer, $h^j_{G_{m,k}} \in \mathbb{R}^{d_1}$ is given by $h^j_{G_{m,k}} = \mathbf{A}_{\mathcal{G'}_{m,k}}\text{LN}[\tanh(T^j_{C_{m,k}}+h^{j-1}_{m,k})]$. Two consecutive GraphSAGE layers are used to aggregate neighborhood information up to two-hop distances. Therefore, for $M$ lines, the embedding is $\underline{h^2_{G_{k}}} \in \mathbb{R}^{M\times d_1}$.

On the other hand, another copy of the embedding, $\underline{h^0_k}$ from the \textit{Line Embedding Layer Block}, is fed to the \textit{Self-Residual Layer Block}, consisting of two self-residual layers, as discussed in Section~IV.~A, to obtain the \textcolor{black}{embedding} $\underline{h^2_{S_k}}\in \mathbb{R}^{M\times d_1}$ in parallel.

The embeddings from both the \textit{Residual GraphSAGE Layer Block}, $\underline{h^2_{G_k}} \in \mathbb{R}^{M\times d_1}$ and \textit{Self-Residual Layer Block}, $\underline{h^2_{S_k}} \in \mathbb{R}^{M\times d_1}$, go through the \textit{Mean-max Pooling Block} \textcolor{black}{individually, giving out two embeddings, $\underline{h^{G}_{pool}}\in \mathbb{R}^{M\times d_1}$ and  $\underline{h^{S}_{pool}}\in \mathbb{R}^{M\times d_1}$, respectively. On one hand, the $\underline{h^{G}_{pool}}$ is combined with the global features, $\underline{\mathbf{Z}_k}$ into an embedding, $[\underline{h^G_{pool}}; \underline{\mathbf{Z}_k}] \in \mathbb{R}^{M\times 2d_1+6}$, and the \textit{Transformed Global Embedding Block}, discussed in Section~IV.~A, converts it into $\underline{h^G_{{T_k}}} \in \mathbb{R}^{M\times d_1}$. On the other hand, the $\underline{h^{S}_{pool}}$ is combined with the global features, $\underline{\mathbf{Z}_k}$ into another embedding, $[\underline{h^S_{pool}}; \underline{\mathbf{Z}_k}] \in \mathbb{R}^{M\times 2d_1+6}$, and the \textit{Transformed Global Embedding Block} converts it into $\underline{h^S_{{T_k}}} \in \mathbb{R}^{M\times d_1}$.}

\textit{Q-value Head and No-action Head Blocks:} The Q-value head takes the positional feature $\underline{\mathbf{E}_\mathcal{G'}}$, the {\it Residual GraphSAGE layer} output $\underline{h^2_{G_k}}$, and the {\it Transformed Global Embedding}, $\underline{h^G_{{T_k}}}$, to determine the Q-values for each of the $M$ lines. On the other hand, the no-action head takes the {\it Transformed Global Embedding Layer} output, $\underline{h^S_{{T_k}}}$, as the only input to calculate a single no-action Q-value. The Q-values from these two heads are concatenated into $M+1$ Q-values, \textcolor{black}{the invalid lines are filtered through the action mask,} and the action at time $k$ is chosen using the $\epsilon$-greedy mechanism as mentioned in the \textit{DQN Value Update Block} in Section~IV.~A. 
The replay buffer, exploration, optimizer, and the updates in the Graph-DQN follow the same mechanism discussed in the \textit{DQN Value Update Block} in Section~IV.~A. 

\subsection{Graph A2C and Graph PPO Models}
Graph A2C and Graph PPO use residual GraphSAGE layers as the actor and critic network alongside the self-residual layers that is used in the graph-agnostic A2C and PPO models. 
Similar to the A2C and PPO models described before, these models are divided into \emph{Actor Network Block} and \emph{Critic Network Blocks}, with additional blocks to capture graph information including \textit{Positional Feature Block} and \textit{Residual GraphSAGE Layer Blocks} in both of the two networks. The details of the modifications made to A2C and PPO models are presented below.

\emph{Actor Network Block:}
The actor network for both A2C and PPO consists of the \textit{Line Embedding Layer Block}, which takes the line-localized features $\underline{\textbf{X}_k}$ and positional feature $\underline{\mathbf{E}_\mathcal{G'}}$ as input and converts them into an embedding, $h^{\text{actor}^0}_k \in \mathbb{R}^{M\times d_1}$ for time step $k$, where $d_1=128$ is considered for the actor network. The embeddings are then sent to the \textit{Self-residual Layer Block} and \textit{Residual GraphSAGE Layer Block} to generate the embeddings, $\underline{h^{\text{actor}^2}_{S_k}} \in \mathbb{R}^{M\times d_1}$ and $\underline{h^{\text{actor}^2}_{G_k}} \in \mathbb{R}^{M\times d_1}$, respectively. 
Two self-residual layers and two GraphSAGE layers are used for the \textit{Self-residual Layer Block} and the \textit{Residual GraphSAGE Layer Block}, respectively, as mentioned in Section~V.~A.

The embeddings $\underline{h^{\text{actor}^2}_{G_k}}$ and $\underline{h^{\text{actor}^2}_{S_k}}$ are sent to two individual \textit{Mean-max Pooling Block} to obtain pooled embedding $\underline{h^{G_{actor}}_{pool}}$ and $\underline{h^{S_{actor}}_{pool}}$ respectively, each with dimension $\mathbb{R}^{M \times 2d_1}$. As discussed in Section~V.~A, the embedding $\underline{h^{G_{actor}}_{pool}}$ is combined with the global features, $\underline{\mathbf{Z}_k}$ into an embedding $[\underline{h^{G_{actor}}_{pool}}; \underline{\mathbf{Z}_k}]\in \mathbb{R}^{M\times 2d_1 +6}$, and the \textit{Transformed Global Embedding Block} converts it into $\underline{h^{\text{actor}}_{G_{T_k}}} \in \mathbb{R}^{M \times d_1}$. On the other hand, the embedding $\underline{h^{G_{actor}}_{pool}}$ is combined with the global features, $\underline{\mathbf{Z}_k}$ into an embedding $[\underline{h^{S_{actor}}_{pool}}; \underline{\mathbf{Z}_k}]\in \mathbb{R}^{M\times 2d_1 +6}$, and the \textit{Transformed Global Embedding Block} converts it into $\underline{h^{\text{actor}}_{S_{T_k}}} \in \mathbb{R}^{M \times d_1}$.

The actor network includes two decision-making heads, referred to as the `line-scoring head' and the `no-action head', which jointly determine the probability distribution over possible actions. The positional feature $\underline{\mathbf{E}_\mathcal{G'}}$, the Residual GraphSAGE output $\underline{h^2_{G_k}}$, and the Transformed Global embedding, $\underline{h^{\text{actor}}_{G_{T_k}}}$ go into the `line scoring head' to determine $M$ logits for $M$ lines. On the other hand, the `no-action head' takes $\underline{h^{\text{actor}}_{S_{T_k}}}$ as the only input to calculate one logit for the no-action. These $M+1$ logits are collected from both the heads, \textcolor{black}{passed through the action mask to filter invalid lines,} and are converted into probabilities using the softmax function to pick the action for the next steps.

\emph{Critic Network Block:}
Similar to the critic network discussed in Section~IV.~B, this critic network takes the line feature vector, $\underline{\mathbf{X}_k}$, and positional features $\underline{\mathbf{E}_\mathcal{G'}}$ as input through the \textit{Line Embedding Layer Block}, combining them into an input of dimension $\mathbb{R}^{M \times 8}$ and converting them into the embedding $\underline{h^{\text{critic}^0}_k} \in \mathbb{R}^{M\times d_2}$ for $M$ lines, where $d_2 = 192$ is the critic network width. Instead of passing the resulting embedding through the Self-residual Layer, the graph-aware critic uses the \textit{Residual GraphSAGE Layer Block} with two GraphSAGE layers to obtain the embedding, $\underline{h^{\text{critic}^2}_{G_k}} \in \mathbb{R}^{M\times d_2}$, which are sent to the \textit{Mean-max Pooling Block} to obtain $\underline{h^{critic}_{pool}}$ and then to the \textit{Transformed Global Embedding Block} to obtain the embedding, $\underline{h^{\text{critic}}_{G_{T_k}}} \in \mathbb{R}^{M \times d_2}$. This embedding is then passed to a \textit{Critic Value Head Block} to generate the critic value function, $V_{\phi}(s_k)$.

The action from the \emph{Actor Network Block} and the critic value generated from the \emph{Critic Network Block} are fed to the \emph{A2C/PPO Policy Update Block}, where each model updates the policy using the update mechanism described in Section~IV.~B. The policy updates from the loss functions for both of the models are then fed back to the actor and critic network blocks to update the gradients and the actor-critic parameters. The hyperparameters used for A2C and PPO are the same as those used for the graph-agnostic A2C and PPO models to ensure a fair comparison among the techniques.

\begin{figure*}[!t]
\centering
\includegraphics[height=5cm,width=14cm]{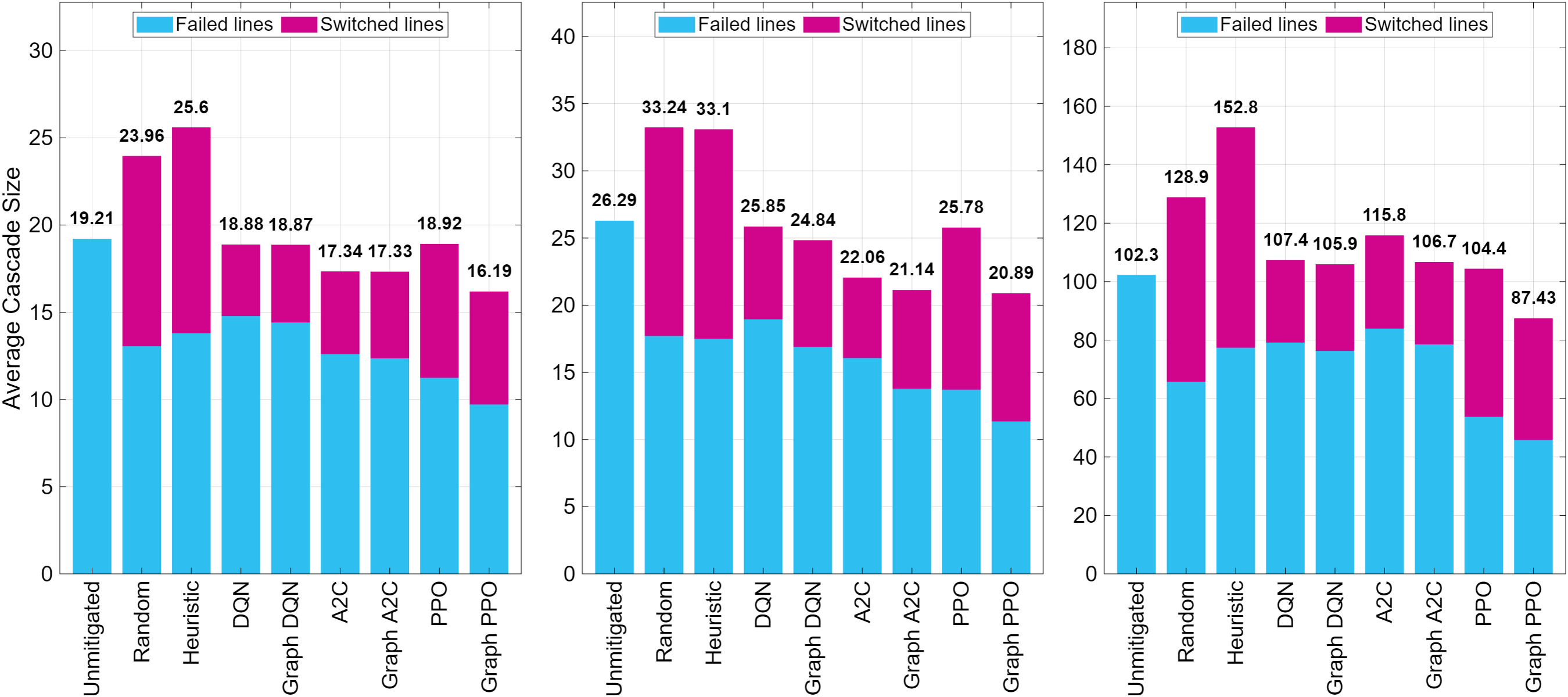}
\par {(a)\qquad\qquad\qquad\qquad\qquad\qquad(b)\qquad\qquad\qquad\qquad\qquad\qquad(c)}
\caption{Average cascade size, including the number of intentionally switched lines and lines that fail due to cascade progression, under different control strategies for the (a) IEEE 39, (b) IEEE 57, and (c) IEEE 118-bus test systems.}
\label{fig:size}
\end{figure*}

\section{Experimental Results and Discussions}

\subsection{Experimental Setup}

In this paper, the cascade mitigation policies are evaluated on three benchmark IEEE test cases: the IEEE 39, 57, and 118 bus systems, consisting of 46, 80, and 186 transmission lines, respectively. MATPOWER \cite{zimmerman2010matpower}, a MATLAB-based tool for power flow calculation, is used to generate 500 cascading scenarios with two random initial failures following the approach in \cite{rahnamay2014stochastic} as reviewed in Section~III.~A.
\textcolor{black}{The DRL agent was trained using 400 distinct cascading failure scenarios and evaluated on 100 previously unseen scenarios.}
All the techniques are trained for the same 50 epochs to ensure consistency. A discount factor of 1 is considered for all the techniques, rewarding each step equally, since the tasks are episodic with finite steps and with confirmed terminations for each scenario. The techniques are implemented in MATLAB to enable efficient interaction with the MATPOWER environment for cascade simulation. At each cascade step, MATPOWER is used to obtain the system states and evaluate the resulting rewards, which are then used to update the learning model.

\begin{figure*}[!t]
\centering
\includegraphics[height=5cm,width=14cm]{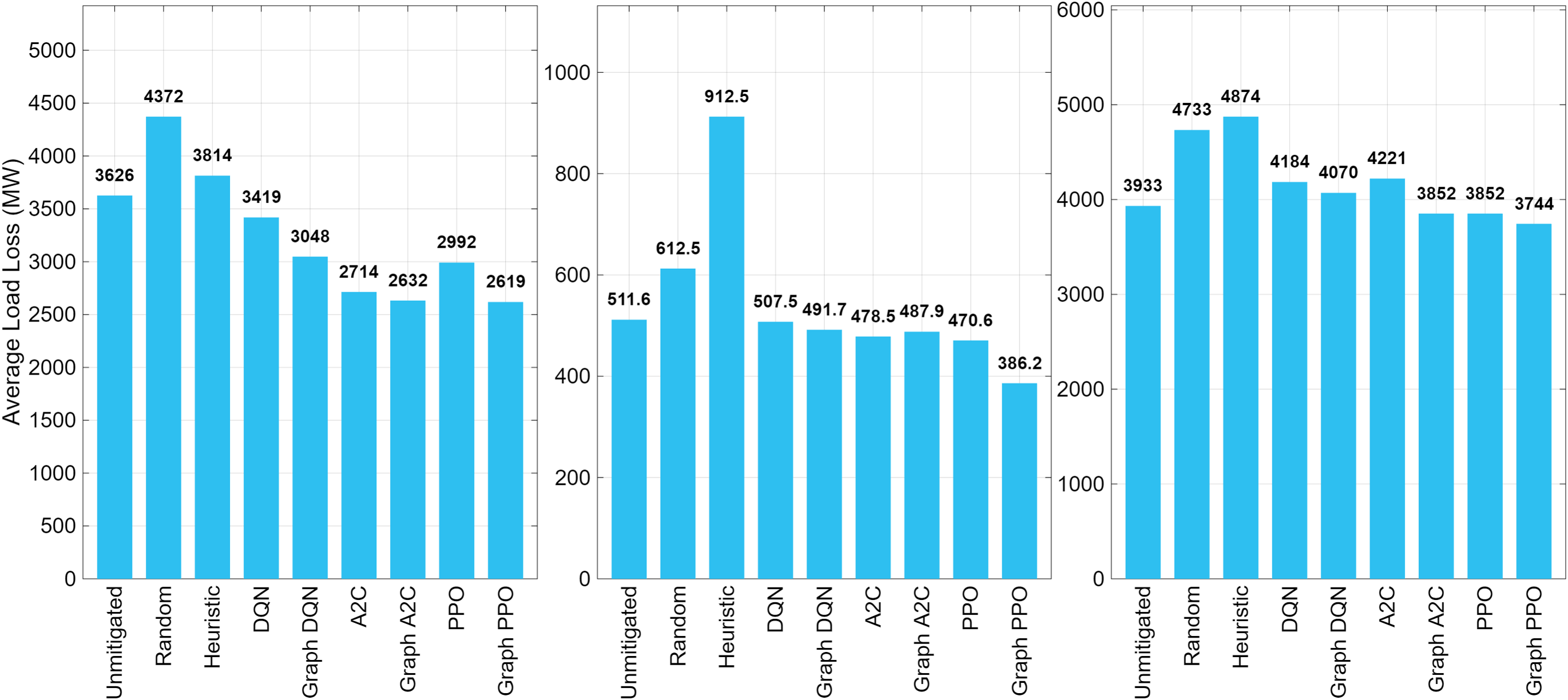}
\par {(a)\qquad\qquad\qquad\qquad\qquad\qquad(b)\qquad\qquad\qquad\qquad\qquad\qquad(c)}
\caption{Average load loss under different control strategies for (a) IEEE 39, (b) IEEE 57, and (c) IEEE 118 test systems.}
\label{fig:loadloss}
\end{figure*}

\begin{figure*}[!t]
\centering
\includegraphics[height=5cm,width=14cm]{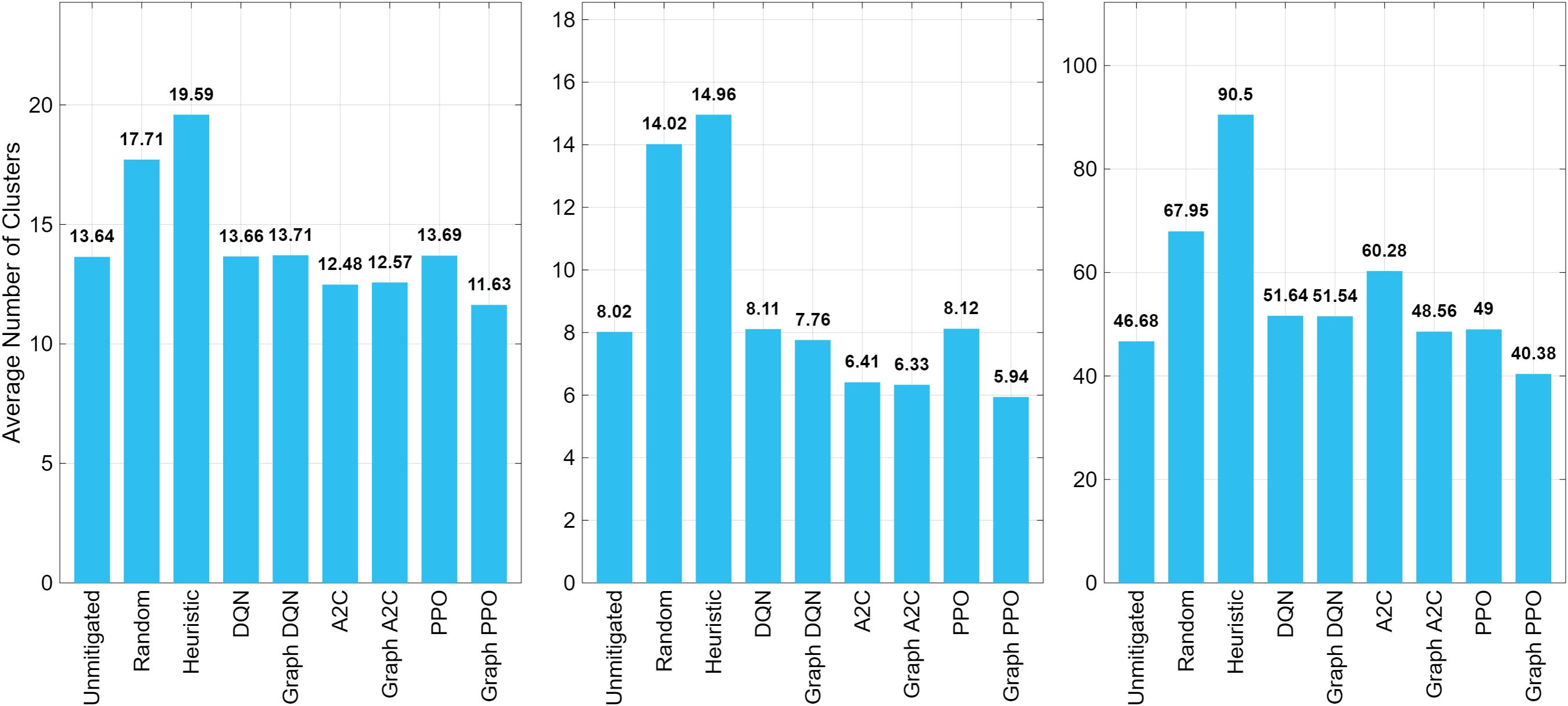}
\par {(a)\qquad\qquad\qquad\qquad\qquad\qquad(b)\qquad\qquad\qquad\qquad\qquad\qquad(c)}
\caption{Average number of clusters formed in the system due to failures or line switches under different control strategies for (a) IEEE 39, (b) IEEE 57, and (c) IEEE 118 test systems.}
\label{fig:cluster}
\end{figure*}

\begin{figure*}[!t]
\centering
\includegraphics[height=4.5cm,width=15cm]{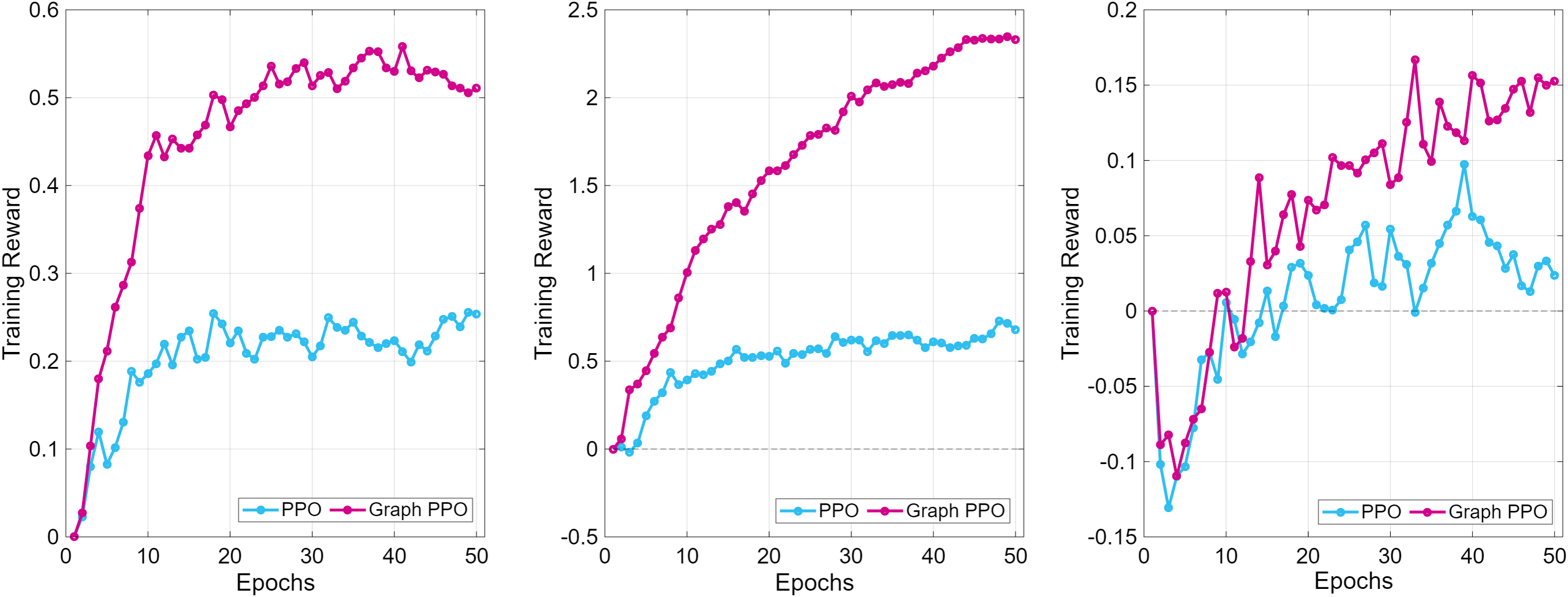}
\par {(a)\qquad\qquad\qquad\qquad\qquad\qquad\quad(b)\quad\qquad\qquad\qquad\qquad\qquad\qquad(c)}
\caption{Evolution of the normalized training rewards for PPO and Graph PPO on the (a) IEEE 39, (b) IEEE 57, and (c) IEEE 118-bus test systems.}
\label{fig:reward}
\end{figure*}

\begin{table*}[]
\centering
\caption{Statistics for the 100 test scenarios for different policies and systems.}
\begin{tabular}{|ccccccccccccc|}
\hline
\multicolumn{13}{|c|}{\textbf{IEEE 39}} \\ \hline
\multicolumn{1}{|c|}{\multirow{2}{*}{\textbf{Policy}}} &
  \multicolumn{4}{c|}{\textbf{Total Cascade Size}} &
  \multicolumn{4}{c|}{\textbf{Total Load Loss}} &
  \multicolumn{4}{c|}{\textbf{Total Number of Clusters}} \\ \cline{2-13} 
\multicolumn{1}{|c|}{} &
  \multicolumn{1}{c|}{\textbf{Min}} &
  \multicolumn{1}{c|}{\textbf{Max}} &
  \multicolumn{1}{c|}{\textbf{Avg}} &
  \multicolumn{1}{c|}{\textbf{Std}} &
  \multicolumn{1}{c|}{\textbf{Min}} &
  \multicolumn{1}{c|}{\textbf{Max}} &
  \multicolumn{1}{c|}{\textbf{Avg}} &
  \multicolumn{1}{c|}{\textbf{Std}} &
  \multicolumn{1}{c|}{\textbf{Min}} &
  \multicolumn{1}{c|}{\textbf{Max}} &
  \multicolumn{1}{c|}{\textbf{Avg}} &
  \textbf{Std} \\ \hline
\multicolumn{1}{|c|}{\textbf{Unmitigated}} &
  \multicolumn{1}{c|}{13} &
  \multicolumn{1}{c|}{22} &
  \multicolumn{1}{c|}{19.21} &
  \multicolumn{1}{c|}{1.91} &
  \multicolumn{1}{c|}{2560.99} &
  \multicolumn{1}{c|}{5019.09} &
  \multicolumn{1}{c|}{3626.27} &
  \multicolumn{1}{c|}{614.60} &
  \multicolumn{1}{c|}{10} &
  \multicolumn{1}{c|}{16} &
  \multicolumn{1}{c|}{13.64} &
  1.39 \\ \hline
\multicolumn{1}{|c|}{\textbf{Random}} &
  \multicolumn{1}{c|}{18} &
  \multicolumn{1}{c|}{30} &
  \multicolumn{1}{c|}{23.96} &
  \multicolumn{1}{c|}{2.70} &
  \multicolumn{1}{c|}{3417.31} &
  \multicolumn{1}{c|}{5412.74} &
  \multicolumn{1}{c|}{4372} &
  \multicolumn{1}{c|}{431.93} &
  \multicolumn{1}{c|}{12} &
  \multicolumn{1}{c|}{23} &
  \multicolumn{1}{c|}{17.71} &
  2.34 \\ \hline
\multicolumn{1}{|c|}{\textbf{Heuristic}} &
  \multicolumn{1}{c|}{20} &
  \multicolumn{1}{c|}{28} &
  \multicolumn{1}{c|}{25.6} &
  \multicolumn{1}{c|}{1.39} &
  \multicolumn{1}{c|}{3015.82} &
  \multicolumn{1}{c|}{4750.49} &
  \multicolumn{1}{c|}{3813.64} &
  \multicolumn{1}{c|}{408.86} &
  \multicolumn{1}{c|}{15} &
  \multicolumn{1}{c|}{22} &
  \multicolumn{1}{c|}{19.59} &
  1.24 \\ \hline
\multicolumn{1}{|c|}{\textbf{DQN}} &
  \multicolumn{1}{c|}{13} &
  \multicolumn{1}{c|}{23} &
  \multicolumn{1}{c|}{18.88} &
  \multicolumn{1}{c|}{1.63} &
  \multicolumn{1}{c|}{2560.99} &
  \multicolumn{1}{c|}{4799.54} &
  \multicolumn{1}{c|}{3418.94} &
  \multicolumn{1}{c|}{522.94} &
  \multicolumn{1}{c|}{10} &
  \multicolumn{1}{c|}{17} &
  \multicolumn{1}{c|}{13.66} &
  1.35 \\ \hline
\multicolumn{1}{|c|}{\textbf{Graph DQN}} &
  \multicolumn{1}{c|}{13} &
  \multicolumn{1}{c|}{22} &
  \multicolumn{1}{c|}{18.87} &
  \multicolumn{1}{c|}{1.39} &
  \multicolumn{1}{c|}{1963.24} &
  \multicolumn{1}{c|}{4745.40} &
  \multicolumn{1}{c|}{3047.97} &
  \multicolumn{1}{c|}{508.14} &
  \multicolumn{1}{c|}{10} &
  \multicolumn{1}{c|}{16} &
  \multicolumn{1}{c|}{13.71} &
  1.10 \\ \hline
\multicolumn{1}{|c|}{\textbf{A2C}} &
  \multicolumn{1}{c|}{14} &
  \multicolumn{1}{c|}{20} &
  \multicolumn{1}{c|}{17.34} &
  \multicolumn{1}{c|}{1.39} &
  \multicolumn{1}{c|}{2139.68} &
  \multicolumn{1}{c|}{4160.53} &
  \multicolumn{1}{c|}{2714.20} &
  \multicolumn{1}{c|}{410.06} &
  \multicolumn{1}{c|}{11} &
  \multicolumn{1}{c|}{14} &
  \multicolumn{1}{c|}{12.48} &
  0.82 \\ \hline
\multicolumn{1}{|c|}
{\textbf{Graph A2C}} &
  \multicolumn{1}{c|}{14} &
  \multicolumn{1}{c|}{21} &
  \multicolumn{1}{c|}{17.33} &
  \multicolumn{1}{c|}{1.58} &
  \multicolumn{1}{c|}{2034.94} &
  \multicolumn{1}{c|}{3801.96} &
  \multicolumn{1}{c|}{2637.03} &
  \multicolumn{1}{c|}{370.64} &
  \multicolumn{1}{c|}{11} &
  \multicolumn{1}{c|}{15} &
  \multicolumn{1}{c|}{12.57} &
  0.92 \\ \hline
\multicolumn{1}{|c|}{\textbf{PPO}} &
  \multicolumn{1}{c|}{15} &
  \multicolumn{1}{c|}{21} &
  \multicolumn{1}{c|}{18.92} &
  \multicolumn{1}{c|}{1.32} &
  \multicolumn{1}{c|}{2140.46} &
  \multicolumn{1}{c|}{4356.75} &
  \multicolumn{1}{c|}{2991.88} &
  \multicolumn{1}{c|}{502.14} &
  \multicolumn{1}{c|}{11} &
  \multicolumn{1}{c|}{16} &
  \multicolumn{1}{c|}{13.69} &
  1.12 \\ \hline
\multicolumn{1}{|c|}{\textbf{Graph PPO}} &
  \multicolumn{1}{c|}{12} &
  \multicolumn{1}{c|}{19} &
  \multicolumn{1}{c|}{\textbf{16.19}} &
  \multicolumn{1}{c|}{1.77} &
  \multicolumn{1}{c|}{1963.24} &
  \multicolumn{1}{c|}{3775.95} &
  \multicolumn{1}{c|}{\textbf{2618.72}} &
  \multicolumn{1}{c|}{405.44} &
  \multicolumn{1}{c|}{10} &
  \multicolumn{1}{c|}{14} &
  \multicolumn{1}{c|}{\textbf{11.63}} &
  1.13 \\ \hline
\multicolumn{13}{|c|}{\textbf{IEEE 57}} \\ \hline
\multicolumn{1}{|c|}{\multirow{2}{*}{\textbf{Policy}}} &
  \multicolumn{4}{c|}{\textbf{Total Size}} &
  \multicolumn{4}{c|}{\textbf{Total Shed}} &
  \multicolumn{4}{c|}{\textbf{Total Cluster}} \\ \cline{2-13} 
\multicolumn{1}{|c|}{} &
  \multicolumn{1}{c|}{\textbf{Min}} &
  \multicolumn{1}{c|}{\textbf{Max}} &
  \multicolumn{1}{c|}{\textbf{Avg}} &
  \multicolumn{1}{c|}{\textbf{Std}} &
  \multicolumn{1}{c|}{\textbf{Min}} &
  \multicolumn{1}{c|}{\textbf{Max}} &
  \multicolumn{1}{c|}{\textbf{Avg}} &
  \multicolumn{1}{c|}{\textbf{Std}} &
  \multicolumn{1}{c|}{\textbf{Min}} &
  \multicolumn{1}{c|}{\textbf{Max}} &
  \multicolumn{1}{c|}{\textbf{Avg}} &
  \textbf{Std} \\ \hline
\multicolumn{1}{|c|}{\textbf{Unmitigated}} &
  \multicolumn{1}{c|}{20} &
  \multicolumn{1}{c|}{30} &
  \multicolumn{1}{c|}{26.29} &
  \multicolumn{1}{c|}{2.16} &
  \multicolumn{1}{c|}{404.51} &
  \multicolumn{1}{c|}{749.68} &
  \multicolumn{1}{c|}{511.60} &
  \multicolumn{1}{c|}{59.65} &
  \multicolumn{1}{c|}{5} &
  \multicolumn{1}{c|}{10} &
  \multicolumn{1}{c|}{8.02} &
  1.20 \\ \hline
\multicolumn{1}{|c|}{\textbf{Random}} &
  \multicolumn{1}{c|}{24} &
  \multicolumn{1}{c|}{42} &
  \multicolumn{1}{c|}{33.24} &
  \multicolumn{1}{c|}{3.78} &
  \multicolumn{1}{c|}{444.74} &
  \multicolumn{1}{c|}{796.41} &
  \multicolumn{1}{c|}{612.49} &
  \multicolumn{1}{c|}{95.73} &
  \multicolumn{1}{c|}{7} &
  \multicolumn{1}{c|}{22} &
  \multicolumn{1}{c|}{14.02} &
  3.05 \\ \hline
\multicolumn{1}{|c|}{\textbf{Heuristic}} &
  \multicolumn{1}{c|}{30} &
  \multicolumn{1}{c|}{38} &
  \multicolumn{1}{c|}{33.1} &
  \multicolumn{1}{c|}{1.85} &
  \multicolumn{1}{c|}{775.36} &
  \multicolumn{1}{c|}{1024.20} &
  \multicolumn{1}{c|}{912.45} &
  \multicolumn{1}{c|}{56.22} &
  \multicolumn{1}{c|}{12} &
  \multicolumn{1}{c|}{18} &
  \multicolumn{1}{c|}{14.96} &
  1.34 \\ \hline
\multicolumn{1}{|c|}{\textbf{DQN}} &
  \multicolumn{1}{c|}{21} &
  \multicolumn{1}{c|}{33} &
  \multicolumn{1}{c|}{25.85} &
  \multicolumn{1}{c|}{2.18} &
  \multicolumn{1}{c|}{383.61} &
  \multicolumn{1}{c|}{659.32} &
  \multicolumn{1}{c|}{507.55} &
  \multicolumn{1}{c|}{56.84} &
  \multicolumn{1}{c|}{5} &
  \multicolumn{1}{c|}{13} &
  \multicolumn{1}{c|}{8.11} &
  1.42 \\ \hline
\multicolumn{1}{|c|}{\textbf{Graph DQN}} &
  \multicolumn{1}{c|}{21} &
  \multicolumn{1}{c|}{30} &
  \multicolumn{1}{c|}{24.84} &
  \multicolumn{1}{c|}{1.64} &
  \multicolumn{1}{c|}{352.90} &
  \multicolumn{1}{c|}{709.58} &
  \multicolumn{1}{c|}{491.65} &
  \multicolumn{1}{c|}{78.45} &
  \multicolumn{1}{c|}{5} &
  \multicolumn{1}{c|}{12} &
  \multicolumn{1}{c|}{7.76} &
  1.22 \\ \hline
\multicolumn{1}{|c|}{\textbf{A2C}} &
  \multicolumn{1}{c|}{19} &
  \multicolumn{1}{c|}{26} &
  \multicolumn{1}{c|}{22.06} &
  \multicolumn{1}{c|}{1.52} &
  \multicolumn{1}{c|}{372.09} &
  \multicolumn{1}{c|}{704.82} &
  \multicolumn{1}{c|}{478.47} &
  \multicolumn{1}{c|}{50.23} &
  \multicolumn{1}{c|}{5} &
  \multicolumn{1}{c|}{8} &
  \multicolumn{1}{c|}{6.41} &
  0.88 \\ \hline
\multicolumn{1}{|c|}
{\textbf{Graph A2C}} &
  \multicolumn{1}{c|}{18} &
  \multicolumn{1}{c|}{24} &
  \multicolumn{1}{c|}{21.14} &
  \multicolumn{1}{c|}{1.26} &
  \multicolumn{1}{c|}{383.89} &
  \multicolumn{1}{c|}{672.02} &
  \multicolumn{1}{c|}{487.92} &
  \multicolumn{1}{c|}{46.47} &
  \multicolumn{1}{c|}{5} &
  \multicolumn{1}{c|}{8} &
  \multicolumn{1}{c|}{6.33} &
  0.85 \\ \hline
\multicolumn{1}{|c|}{\textbf{PPO}} &
  \multicolumn{1}{c|}{22} &
  \multicolumn{1}{c|}{31} &
  \multicolumn{1}{c|}{25.78} &
  \multicolumn{1}{c|}{2.11} &
  \multicolumn{1}{c|}{346.50} &
  \multicolumn{1}{c|}{708.38} &
  \multicolumn{1}{c|}{470.63} &
  \multicolumn{1}{c|}{74.44} &
  \multicolumn{1}{c|}{5} &
  \multicolumn{1}{c|}{12} &
  \multicolumn{1}{c|}{8.12} &
  1.65 \\ \hline
\multicolumn{1}{|c|}{\textbf{Graph PPO}} &
  \multicolumn{1}{c|}{16} &
  \multicolumn{1}{c|}{26} &
  \multicolumn{1}{c|}{\textbf{20.89}} &
  \multicolumn{1}{c|}{1.69} &
  \multicolumn{1}{c|}{320.91} &
  \multicolumn{1}{c|}{570.74} &
  \multicolumn{1}{c|}{\textbf{386.17}} &
  \multicolumn{1}{c|}{64.22} &
  \multicolumn{1}{c|}{4} &
  \multicolumn{1}{c|}{11} &
  \multicolumn{1}{c|}{\textbf{5.94}} &
  1.25 \\ \hline
\multicolumn{13}{|c|}{\textbf{IEEE 118}} \\ \hline
\multicolumn{1}{|c|}{\multirow{2}{*}{\textbf{Policy}}} &
  \multicolumn{4}{c|}{\textbf{Total Size}} &
  \multicolumn{4}{c|}{\textbf{Total Shed}} &
  \multicolumn{4}{c|}{\textbf{Total Cluster}} \\ \cline{2-13} 
\multicolumn{1}{|c|}{} &
  \multicolumn{1}{c|}{\textbf{Min}} &
  \multicolumn{1}{c|}{\textbf{Max}} &
  \multicolumn{1}{c|}{\textbf{Avg}} &
  \multicolumn{1}{c|}{\textbf{Std}} &
  \multicolumn{1}{c|}{\textbf{Min}} &
  \multicolumn{1}{c|}{\textbf{Max}} &
  \multicolumn{1}{c|}{\textbf{Avg}} &
  \multicolumn{1}{c|}{\textbf{Std}} &
  \multicolumn{1}{c|}{\textbf{Min}} &
  \multicolumn{1}{c|}{\textbf{Max}} &
  \multicolumn{1}{c|}{\textbf{Avg}} &
  \textbf{Std} \\ \hline
\multicolumn{1}{|c|}{\textbf{Unmitigated}} &
  \multicolumn{1}{c|}{97} &
  \multicolumn{1}{c|}{108} &
  \multicolumn{1}{c|}{102.32} &
  \multicolumn{1}{c|}{2.22} &
  \multicolumn{1}{c|}{3588.28} &
  \multicolumn{1}{c|}{4247.48} &
  \multicolumn{1}{c|}{3932.73} &
  \multicolumn{1}{c|}{126.17} &
  \multicolumn{1}{c|}{43} &
  \multicolumn{1}{c|}{50} &
  \multicolumn{1}{c|}{46.68} &
  1.74 \\ \hline
\multicolumn{1}{|c|}{\textbf{Random}} &
  \multicolumn{1}{c|}{116} &
  \multicolumn{1}{c|}{141} &
  \multicolumn{1}{c|}{128.90} &
  \multicolumn{1}{c|}{4.85} &
  \multicolumn{1}{c|}{4410.75} &
  \multicolumn{1}{c|}{5079.34} &
  \multicolumn{1}{c|}{4733.18} &
  \multicolumn{1}{c|}{153.28} &
  \multicolumn{1}{c|}{57} &
  \multicolumn{1}{c|}{80} &
  \multicolumn{1}{c|}{67.95} &
  3.91 \\ \hline
\multicolumn{1}{|c|}{\textbf{Heuristic}} &
  \multicolumn{1}{c|}{140} &
  \multicolumn{1}{c|}{158} &
  \multicolumn{1}{c|}{152.8} &
  \multicolumn{1}{c|}{2.98} &
  \multicolumn{1}{c|}{4367.56} &
  \multicolumn{1}{c|}{5007.45} &
  \multicolumn{1}{c|}{4873.81} &
  \multicolumn{1}{c|}{84.80} &
  \multicolumn{1}{c|}{80} &
  \multicolumn{1}{c|}{95} &
  \multicolumn{1}{c|}{90.50} &
  2.33 \\ \hline
\multicolumn{1}{|c|}{\textbf{DQN}} &
  \multicolumn{1}{c|}{100} &
  \multicolumn{1}{c|}{115} &
  \multicolumn{1}{c|}{107.37} &
  \multicolumn{1}{c|}{3.60} &
  \multicolumn{1}{c|}{3695.94} &
  \multicolumn{1}{c|}{4611.13} &
  \multicolumn{1}{c|}{4184.41} &
  \multicolumn{1}{c|}{154.58} &
  \multicolumn{1}{c|}{47} &
  \multicolumn{1}{c|}{57} &
  \multicolumn{1}{c|}{51.64} &
  2.46 \\ \hline
\multicolumn{1}{|c|}{\textbf{Graph DQN}} &
  \multicolumn{1}{c|}{95} &
  \multicolumn{1}{c|}{115} &
  \multicolumn{1}{c|}{105.93} &
  \multicolumn{1}{c|}{3.45} &
  \multicolumn{1}{c|}{3739.50} &
  \multicolumn{1}{c|}{4424.12} &
  \multicolumn{1}{c|}{4070.43} &
  \multicolumn{1}{c|}{145.56} &
  \multicolumn{1}{c|}{45} &
  \multicolumn{1}{c|}{58} &
  \multicolumn{1}{c|}{51.54} &
  2.51 \\ \hline
\multicolumn{1}{|c|}{\textbf{A2C}} &
  \multicolumn{1}{c|}{109} &
  \multicolumn{1}{c|}{122} &
  \multicolumn{1}{c|}{115.84} &
  \multicolumn{1}{c|}{2.31} &
  \multicolumn{1}{c|}{3856.34} &
  \multicolumn{1}{c|}{4617.27} &
  \multicolumn{1}{c|}{4221.12} &
  \multicolumn{1}{c|}{119.49} &
  \multicolumn{1}{c|}{56} &
  \multicolumn{1}{c|}{65} &
  \multicolumn{1}{c|}{60.28} &
  1.59 \\ \hline
\multicolumn{1}{|c|}
{\textbf{Graph A2C}} &
  \multicolumn{1}{c|}{96} &
  \multicolumn{1}{c|}{111} &
  \multicolumn{1}{c|}{106.73} &
  \multicolumn{1}{c|}{2.59} &
  \multicolumn{1}{c|}{3651.01} &
  \multicolumn{1}{c|}{4164.40} &
  \multicolumn{1}{c|}{3851.87} &
  \multicolumn{1}{c|}{105.07} &
  \multicolumn{1}{c|}{43} &
  \multicolumn{1}{c|}{52} &
  \multicolumn{1}{c|}{48.56} &
  1.88 \\ \hline
\multicolumn{1}{|c|}{\textbf{PPO}} &
  \multicolumn{1}{c|}{97} &
  \multicolumn{1}{c|}{113} &
  \multicolumn{1}{c|}{104.42} &
  \multicolumn{1}{c|}{2.79} &
  \multicolumn{1}{c|}{3508.96} &
  \multicolumn{1}{c|}{4199.70} &
  \multicolumn{1}{c|}{3851.91} &
  \multicolumn{1}{c|}{129.65} &
  \multicolumn{1}{c|}{42} &
  \multicolumn{1}{c|}{55} &
  \multicolumn{1}{c|}{49} &
  2.79 \\ \hline
\multicolumn{1}{|c|}{\textbf{Graph PPO}} &
  \multicolumn{1}{c|}{82} &
  \multicolumn{1}{c|}{96} &
  \multicolumn{1}{c|}{\textbf{87.43}} &
  \multicolumn{1}{c|}{2.58} &
  \multicolumn{1}{c|}{3516.76} &
  \multicolumn{1}{c|}{4107.70} &
  \multicolumn{1}{c|}{\textbf{3743.83}} &
  \multicolumn{1}{c|}{124.15} &
  \multicolumn{1}{c|}{36} &
  \multicolumn{1}{c|}{45} &
  \multicolumn{1}{c|}{\textbf{40.38}} &
  1.55 \\ \hline
\end{tabular}
\end{table*}

In addition to the proposed DRL approaches, three benchmarks are considered as baselines to facilitate a comprehensive performance comparison. The additional benchmarks include the unmitigated cascade with no control (representing the natural progress of cascading failures in the absence of the mitigation mechanisms), a random line switching policy, and a line switching policy based on a Line Outage Distribution Factor (LODF)-based heuristic.
The random switching policy is designed to randomly switch lines or select no-action options at each step of the cascading failures based on a uniform distribution.
The LODF-based heuristic policy calculates and updates the LODF values of the lines as discussed in \cite{LODF-heuristic} after each failure or line switching and selects the line with the minimum LODF as the next line switching candidate. 
This heuristic-based policy is designed with the goal of reducing the risk of further failures by removing the least impactful lines guided by the LODF. All baseline methods and the proposed DRL-based policies are evaluated and compared in terms of cascade mitigation performance, using the total cascade size, the number of electrical islands (clusters) formed, and the amount of load loss as the primary performance metrics.
 
All experiments are run on MATLAB 2026a with parallel computing capability with 20 workers, on a Dell workstation equipped with an Intel(R) Core Ultra 7 265 processor (20 cores, 2.40 GHz), 32 GB of RAM, and an NVIDIA GeForce RTX 4070 Ti SUPER GPU.

\subsection{Results and Discussion}

The performance of the DRL-based methods and baseline approaches is evaluated and discussed in this section.
Figure 2 presents the average cascade size, including the average number of lines intentionally switched as part of the control strategy (purple) and those that fail due to the natural progression of the cascade (cyan) for different techniques. It can be observed from the results that all graph-aware models demonstrate improved performance over their graph-agnostic counterparts. Among these models, Graph PPO consistently shows improvement over the unmitigated baseline, indicating that the line switching chosen by the Graph PPO agents effectively steers the cascade towards earlier termination. Specifically, Graph PPO learns a better mitigation strategy, reducing cascade failures by 15.72\%, 20.54\%, and 14.54\% for the IEEE 39, 57, and 118, compared to the unmitigated cascade baseline, respectively. The improvements in the cascade size are accompanied by an improvement in the served load, which is indicated by the decrease in load loss in the DRL-based techniques, as shown in Figure 3. From Figure 3, it can be observed that Graph PPO shows load loss reductions of 27.77\%, 24.51\%, and 4.81\%  for IEEE 39, 57, and 118 systems compared to the unmitigated cascading failure, respectively. This result indicates that the tradeoff management between cascade size and load loss may be more difficult in larger systems. 
As shown in Figure~4, the number of grid clusters resulting from line failures or intentional line switching is reduced by 14.74\%, 25.94\%, and 13.50\% for the IEEE 39, 57, and 118 bus systems, respectively, compared to the corresponding unmitigated cases. 
\textcolor{black}{Table I} summarizes the minimum, maximum, mean, and standard deviation of the metrics including the total cascade size, load loss, and number of clusters for all the techniques and test systems. The best-performing values for each metric and system are highlighted in bold. Consistent with the previous results, Graph PPO notably outperforms the unmitigated, random, and heuristic baselines, as well as the other DRL methods, across all evaluated metrics.

The improvements of Graph PPO can be attributed to the combination of the graph architecture and the PPO update mechanism. Figure~5 shows the learning mechanism of Graph PPO against PPO for all the systems through cumulative rewards. 
Although for the first few epochs, Graph PPO follows the self-residual layer-based PPO, the residual GraphSAGE layer eventually navigates the values of learnable parameters in later epochs to manage how topological information can aid the model, accompanied by residual connections and normalizations ensuring stability and smoothness. Furthermore, PPO clipping and multiple training epochs over the minibatch contribute to the improvement compared to A2C models. As observed from Figures 2, 3, and 4, Graph PPO outperforms Graph A2C in all three metrics for all the systems, indicating that with a similar graph architecture, the PPO policy updates further help to learn better remedial actions compared to A2C policy updates for the same cascade cases.

The training time for each of the DRL learning models is important to analyze, as DRL training is inherently time-consuming. Table 2 summarizes their training time in hours for each DRL technique for different systems. The on-policy A2C and PPO techniques take significantly less time to train compared to the off-policy DQN due to their use of the replay buffer.
Graph PPO also achieves shorter training times, particularly on the IEEE 57 and 118 bus systems, while delivering performance improvements across all evaluated metrics compared to the other graph-aware approaches. This makes Graph PPO a promising candidate for the emergency line switching mechanisms during an ongoing cascade. 

\begin{table}[]
\caption{Training time (in hours) for each of the DRL models for different test systems.}
\centering
\label{tab:my-table}
\begin{tabular}{|c|c|c|c|}
\hline
\textbf{Model} & \textbf{IEEE 39} & \textbf{IEEE 57} & \textbf{IEEE 118} \\ \hline
DQN            & 1.40             & 2.09             & 23.98             \\ \hline
Graph DQN      & 1.18             & 1.95             & 22.99             \\ \hline
A2C            & 0.78             & 0.59             & 15.01             \\ \hline
Graph A2C            & 0.80             & 0.41             & 13.65             \\ \hline
PPO            & 0.88             & 0.57             & 10.34              \\ \hline
Graph PPO      & 0.93             & 0.55             & 10.27              \\ \hline
\end{tabular}
\end{table}

\section{Conclusion} 
This work formulated adaptive transmission line switching for cascade mitigation in power systems as a DRL problem. Three graph-aware DRL frameworks, namely Graph-DQN, Graph-A2C, and Graph-PPO, were developed to capture power system topology through graph while incorporating global system-state information. The proposed frameworks were evaluated on the IEEE 39, 57, and 118 bus systems and benchmarked against the unmitigated case, and random, and heuristic-based control strategies, as well as their graph-agnostic counterparts. The results demonstrate the effectiveness of incorporating graph information into DRL-based cascade mitigation frameworks. In particular, it was observed that Graph PPO outperformed the evaluated techniques in reducing cascade size, load loss, and the number of clusters formed in the system, while also having shorter training times. The results highlight the importance of incorporating graph information in DRL-based techniques for cascade mitigation in power systems. Future work will explore the joint optimization of multiple corrective control actions, particularly transmission line switching and load shedding, as a promising avenue for enhancing cascade mitigation and load preservation within graph-aware DRL frameworks.

\section*{Acknowledgment}
This material is based upon work supported by the National Science Foundation under Grant No.~2238658.

\ifCLASSOPTIONcaptionsoff
  \newpage
\fi

\vfill

\end{document}